# Water Boiling inside Carbon Nanotubes: Towards Efficient Drug Release

Vitaly V. Chaban and Oleg V. Prezhdo[*]

Department of Chemistry, University of Rochester, Rochester, NY 14627, USA

We show using molecular dynamics simulation that spatial confinement of water inside carbon nanotubes (CNT) substantially increases its boiling temperature and that a small temperature growth above the boiling point dramatically raises the inside pressure. Capillary theory successfully predicts the boiling point elevation down to 2 nm, below which large deviations between the theory and atomistic simulation take place. Water behaves qualitatively different inside narrow CNTs, exhibiting transition into an unusual phase, where pressure is gas-like and grows linearly with temperature, while the diffusion constant is temperature-independent. Precise control over boiling by CNT diameter, together with the rapid growth of inside pressure above the boiling point, suggests a novel drug delivery protocol. Polar drug molecules are packaged inside CNTs; the latter are delivered into living tissues and heated by laser. Solvent boiling destroys CNT capping agents and releases the drug.

[*] Corresponding author. E-mail: vvchaban@gmail.com; v.chaban@rochester.edu

The unusual behavior of liquids confined inside carbon nanotubes (CNTs) creates diverse possibilities for new devices, including isotope separation columns,[1] nano-fluidic,[2, 3] electrophoretic[4] and thermophoretic[5] channels, membranes,[6, 7] nanoscale chemical[8, 9] and electrochemical cells,[10] high power capacitors[11] and batteries,[12] DNA sequencing[13] and therapeutic[14-16] tools. Numerous investigations of confined liquids have been reported in the last decade, with a large fraction of the efforts dedicated to the properties of confined water. Water enters[17] and diffuses through CNTs[18, 19] in a concerted manner, forming hydrogen-bonded molecular files. Confinement within the hydrophobic environment of a CNT softens the dynamics of water molecules,[20] and as a result, the viscosity of water decreases rapidly as the tube radius is reduced or when the flow is increased.[21] Other solvents exhibit less dramatic changes upon confinement.[11, 22]

Experimental and theoretical studies indicate that fluids inside nanopores exhibit a variety of states that differ greatly from those of bulk liquids. Confined at the nanoscale, water forms ice-like structures even under ambient conditions.[23-26] Solute molecules tend to aggregate inside CNTs,[16] and the formation of solute clusters is essential for CNT filling.[27] Metals such as gallium can remain liquid under confinement to temperatures lower than normal and tend to crystalize into unusual phases.[28] Iodine requires a critical nanospace size in order to crystallize. Below the critical size, the iodine structure strongly depends on the CNT diameter.[29] Interaction with CNT can produce new types of materials. For instance, polymeric nitrogen becomes stable at ambient conditions inside CNTs.[30] Similarly, weakly hydrophobic interactions increase stability of peptide helical structures inside sufficiently narrow CNTs, which can be viewed as mimics of the ribosome tunnel.[31]

On the fundamental side, the focus was primarily on transitions between liquid and solid phases. Normally, freezing of a liquid occurs through a first-order phase transition. Under confinement, however, freezing can take place by means of both first-order and continuous phase transitions, and the first-order and continuous transition lines should meet at a certain point.[32, 33] Liquid/vapor coexistence was investigated for n-alkanes confined inside CNTs. The two phases merge at the critical point, at which, relative to unconfined alkanes, temperature decreases and density increases.[34] Generally, the above phenomena can be understood as a result of the entropic effect due to spatial confinement and competition between fluid-fluid and fluid-surface intermolecular interactions.

CNTs draw significant attention as a promising candidate for therapeutic applications.[14, 35] The concerns raised about CNT toxicity[36, 37] apply primarily to longer tubes that cannot be handled by immune cells. The unique combination of CNTs' mechanical and optical properties forms the basis for their use as drug delivery systems. The hydrophobic properties allow CNTs to penetrate through cell membranes and reach those parts of an organism that are not accessible to polar substances. Hence, a polar drug molecule can be encapsulated by a CNT or attached to it, and then delivered inside the organism through lipid bilayers and other hydrophobic tissues.[15, 38-40] In order to release the drug, mechanisms involving biodegradable capping agents,[41] ferromagnets[42] and smaller diameter CNTs[43] have been proposed. Alternatively, non-invasive activation of CNTs can be achieved by near-infrared light, which passes through living tissues and is absorbed by CNTs.[44] The energy of the light rapidly transfers from electrons to phonons[45] and then to the surrounding medium.[46, 47] It has been shown experimentally that this energy can be used to destroy malignant cells[48, 49] or to release drug molecules.[50-52]

The current work investigates the fundamental aspects of water boiling under confinement and shows that the strong dependence of the boiling process on CNT diameter and infrared radiation energy can be used to achieve precise control over the release of an encapsulated drug. Using classical molecular dynamics (MD) simulation, we establish that confinement greatly affects the liquid/vapor phase transition temperature and that pressures of several hundred atmospheres can be achieved easily by a small temperature growth above the phase transition. We find that the capillary theory is valid for CNTs with diameters of 2nm and higher and that large deviations take place for more narrow CNTs. Narrow CNTs alter dramatically not only the ice/liquid phase transition[32, 33] but also the boiling process. Rapid growth of pressure inside CNTs at temperatures above the boiling point, accompanied by a strong dependence of boiling temperature on CNT diameter, leads to a novel drug release proposal. Near-infrared radiation heats CNTs and drives local vaporization of the confined solvent. The evaporated liquid, still located inside the CNT, generates high pressure on the CNT edges, destroys capping agents, and pushes drug molecules outside.

**RESULTS AND DISCUSSION**

The study focused on four CNT/water systems, see Table 1. The CNT diameters ranged from under 1nm to nearly 3nm. Figure 1 shows one of the multi-wall CNT systems used in the simulations of confined water boiling. Details of the dynamics simulations are described in the Methods section. In addition to the CNT/water systems, we investigated the boiling temperature of bulk water, system I in Table 1.

**Boiling of bulk water.** Prior to studying the boiling of water confined inside CNTs, we needed to establish that classical MD can reliably reproduce the boiling temperature of bulk water. We tested several classical water models and found that the TIP4P force-field does the best job in reproducing the boiling point of water. In order to obtain the boiling temperature, we plotted the saturated vapor pressure vs. temperature, Figure 2a, derived from the MD simulation of the liquid/vapor interface, Figure 2b. The original periodic liquid water system, system I in Table 1, was expanded in z-direction. As a result, the bulk water system was transformed into a thin film of water centered in the simulation box and surrounded by several nanometers of vacuum. The volume of the simulation cell was fixed during the subsequent simulation. The system was allowed to equilibrate at several temperatures. A fraction of water molecules from the thin film evaporated, creating saturated water vapor above liquid water. According to the standard definition, the saturated vapor pressure in this model system corresponds to the pressure acting along the z-direction of the simulation cell that is perpendicular to the water film. In turn, the boiling temperature at atmospheric pressure is the temperature for which the pressure of the saturated vapor reaches 1 atm.

The boiling temperature of bulk water obtained in our simulation using the TIP4P water model is about 370 K, Figure 2a. At this temperature, the saturated vapor pressure reaches 1 atm, and the slope of the pressure vs. temperature curve sharply increases. The calculated boiling temperature is quite close to the experimental value of 373K. This is quite encouraging given such a simple classical model of water. The result is in good agreement with other literature data[53, 54] for TIP4P. Figure 2b illustrates the coexistence of liquid water and water vapor in our model system at different temperatures. It shows the local density profile perpendicular to the liquid water film. At temperatures between 270 and 350 K, the vapor density is much less than 1 kg/m$^3$, whereas the density of the water film is about 1 000 kg/m$^3$, equal to that of bulk water. The saturated vapor pressure is small at these temperatures, Figure 2a. The phase transition occurs around 370K. At temperatures starting from 390 K, the vapor density sharply increases, Figure 2b, indicating that the liquid is boiling and rapidly evaporating. These results show that the TIP4P water model successfully reproduces the boiling temperature of bulk water. Therefore, one can apply this model for studying the boiling process under spatial confinement.

**Liquid/vapor phase transition under spatial confinement.** The dependence of water pressure on temperature is shown in Figure 3 for the four CNT systems. At low temperatures, the pressure exerted by confined water along the CNT axis is negligible. As the systems are heated, the pressure rapidly rises once the temperature reaches the liquid/vapor phase transition. Elevation of the boiling temperature due to spatial confinement correlates inversely with the nanotube diameter. The boiling point elevation ranges from 15 K for the (21,21) CNT to 250 K for the (6,6) CNT. Above the phase transition temperature, the pressure grows linearly with temperature. The slope of the pressure vs. temperature line is equal to 19 bar/K for the three largest CNTs. The slope drops to 13 bar/K for the smallest (6,6) CNT, indicating that the liquid/vapor phase transition of water inside the (6,6) tube is different from the phase transition in the larger tubes. Further analysis of the (6,6) system is given below while considering the structure and diffusion coefficient of confined water.

**Capillary theory of liquid/vapor phase transition.** The boiling point elevation of a fluid inside CNTs can be predicted using the framework of capillary theory. Assuming that liquids adsorbed by porous bodies are held up by capillary forces, the lowering of vapor pressure at the meniscus is given by the well-known Kelvin equation. The elevation of the boiling point for a given vapor pressure can be obtained using the Clausius-Clapeyron relation, provided that at least one (T, P) point on the evaporation curve is known. The results of such calculations are presented in Figure 4.

The capillary theory and MD calculations agree very well for the two larger tubes, (21,21) and (16,16), with the inner diameters of 2.55 and 1.87 nm, respectively. When the CNT diameter is decreased significantly below 2nm, capillary theory fails completely, as exemplified by the (11,11) and (6,6) CNTs with the inner diameters of 1.19 and 0.52 nm. Capillary theory greatly underestimates the MD boiling point. Interestingly, the boiling temperatures of the (11,11) and (6,6) systems obtained from MD correspond to the capillary theory estimates derived using the diameters of 0.70 and 0.27 nm, respectively. These are 1.7 and 1.9 times smaller than the true CNT diameters.

The (6,6) CNT is known to encapsulate only a single file of water molecules [55]. Both transport and structural properties of such water chains differ significantly from the properties of the bulk phase. In turn, the (11,11) CNT contains three solvent layers along its radial direction [55]. Two layers are in direct proximity with the CNT walls, and the spatial confinement effect is still strong. It should be noted that the values of the surface tension and heat of vaporization used in the capillary theory equations are taken for bulk water. The deviations of capillary theory from the MD simulations are associated with changes in these values when they are taken to the molecular level. The capillary theory neglects to account for the atomistic details of the CNT-water interactions, which play an increasingly important role as the number of confined solvent layers approaches one.

**Structure of confined gaseous and liquid water.** The local density of confined water computed along the CNT axis is shown in Figure 5 for the smallest and largest CNTs. Each panel contains three lines corresponding to 300 K, a temperature slightly above the phase transition and a high temperature. The data for the two CNTs of intermediate diameters are qualitatively similar to the plots shown for the (21,21) CNT. At 300 K, the confined water is more ordered than above the phase transition, and the local density exhibits sharper peaks.

The (6,6) CNT presents a special case. At room temperature, the single file of water molecules inside the (6,6) CNT forms the so-called ice nanotube. This is evidenced by the extreme periodicity of the local density along the tube axis, top panel of Figure 5. Remarkably, some periodicity still remains above the phase transition point, even though the amplitude of the density oscillation is greatly decreased. Also quite remarkably, the density distribution near the boiling point at the pressure of 1 atm is very similar to the distribution at the pressure that is 500 times higher. These observations, supported further by the temperature dependence of the diffusion coefficient discussed below, lead us to conclude that the phase transition of the single-file water chain inside the (6,6) CNT should not be regarded as a liquid/vapor phase transition. More appropriately, it should be viewed as a transition from an ice-like phase to a pseudo-gas phase, in which pressure grows linearly with temperature, Figure 3, as expected for a gas, but in which the translational mobility of water molecules is greatly restricted. The high density of confined water helps to maintain strong hydrogen bonding along the file of water molecules, inhibiting translational motion. The pressure is created by vibrations of water molecules with respect to each other rather than by translations, as in regular gas phase.

**Diffusion of confined water above the liquid/vapor phase transition.** Figure 6 shows the translational diffusion constants of confined water at temperatures when the pressure of the confined medium exceeds 1 bar. The temperature dependence of the diffusion coefficient is different for the (6,6) CNT compared to the larger tubes. The diffusion constant of water inside the smallest tube is about three times lower than that of bulk water at 298 K ($2.5 \times 10^{-9}$ m$^2$/s). Moreover, the diffusion coefficient shows no temperature dependence. In contrast, the temperature dependence of the diffusion coefficient of water inside the (11,11), (16,16) and (21,21) CNTs is linear within the considered temperature intervals. The corresponding correlation coefficients exceed 0.99. The linear temperature dependence of the diffusion coefficient follows from the kinetic theory of gases and differs from the Arrhenius dependence that arises in Vogel-Tamman-Fulcher theory. For instance, Arrhenius dependence is observed for a polar, π-conjugated, organic drug molecule in liquid water confined inside CNTs.[16] Above the boiling point, the diffusion coefficient of water inside the (11,11), (16,16) and (21,21) CNTs is only 3-4 times higher than the diffusion coefficient of bulk water at 300K. The difference is extremely small compared to the 1-2 orders of magnitude difference between the diffusion coefficients of vapor and liquid water without confinement. The small values of the diffusion coefficients of confined water above the phase transition temperature support the facts that translational motions are greatly restricted in the considered systems and that the internal pressure is generated by vibrational motions of water molecules with respect to each other.

Considering the temperature dependence of the diffusion coefficient of water confined inside the (6,6) CNT, Figure 6, one can speculate that generally under spatial confinement, the diffusion coefficient should reach a plateau at a very high temperature. This temperature value should depend on the pore size and should strongly correlate with the fluid density. The phase transition temperature is much higher for the (6,6) CNT compared to the other tubes, and the plateau is already reached at temperatures below the phase transition. For larger tubes the plateau temperature is should be significantly higher than the phase transition temperature.

**Implications of boiling under confinement for drug delivery.** The results of our MD simulations of water boiling inside CNTs indicate that heating of confined systems by 20-30 K above their boiling points creates a huge pressure increase from below 1 atm to 500 atm, Figure 3. The corresponding pressure increase for bulk water is much more modest, Figure 2. The enormous growth of pressure inside CNTs following the phase transition can be utilized in drug delivery applications discussed in the introduction. High pressures generated inside CNTs upon modest heating can help to remove CNT capping agents and facilitate drug release from CNTs into the surrounding medium.

The reported study shows that rather large absolute temperatures are required in order to achieve high pressure in the case of water, especially for smaller tubes. Although local overheating *per se* should not be fatal to the living organism due to the size difference between nanotubes and living cells, it is desirable and more efficient to consider other solvents that evaporate at lower temperatures. CNT heating to high temperatures will require substantial laser power, and long-time exposure to high-power heating will result in accumulation of excess energy that can be potentially dangerous for living cells.

Our study shows that the boiling temperature of a confined solvent can be controlled precisely by the CNT diameter. Therefore, by tuning the nanotube size in conjunction with an appropriate solvent choice, one can achieve the desired boiling temperature. The boiling point should be above the temperature of the living body, since otherwise drug molecules will leave CNT before it is delivered to the target cell. Since a 20-30 K increase in temperature above the boiling point is already sufficient to generate very high pressures inside CNT, an ideal boiling temperature of confined solvent should be around 315-330 K.

Consider carbon dioxide ($CO_2$) that is inexpensive and non-toxic. At atmospheric pressure, $CO_2$ evaporates at 194.5 K. The capillary theory predicts that $CO_2$ will boil at 203, 206, 214 and 249 K inside the (21,21), (16,16), (11,11) and (6,6) CNTs, respectively. If one assumes that in small diameter CNTs the linear O=C=O molecules are likely to be parallel to the CNT axis, then the number of $CO_2$ layers in the direction perpendicular to the CNT axis should be very close to the case of water. Hence, the deviations of the capillary theory from the MD simulation derived for water can be approximately transferred to carbon dioxide in order to correct the boiling temperatures for the (6,6) and (11,11) CNTs. With these corrections, we obtain 321 and 229 K, respectively. The first value lies within the ideal temperature interval. Unfortunately, the (6,6) CNT is too small to encapsulate any real drug molecule.

**Drug delivery proposal.** The above analysis indicates that the boiling temperature of water is too high, while the boiling temperature of $CO_2$ is too low to achieve the desired control over the drug release within the ideal temperature range. Therefore, one is naturally led to consider a mixture of these two solvents. Such mixture is known as carbonated water, and the equimolar mixture of $H_2O$ and $CO_2$ is equivalent to carbonic acid with the empirical formula $H_2CO_3$. Carbonic acid is an unstable compound that exists only in the liquid phase. Hence, it has no boiling point. When heated, carbonated water releases $CO_2$ in a process akin to boiling. Upon laser heating of CNTs, carbonated water confined inside the tubes will emit $CO_2$ molecules, which will break their bonds with water molecules and generate substantial pressure. The pressure will act on the nanotube ends, destroying the capping agents and pushing drug molecules outside of CNTs, Figure 7. The minute amounts of carbonic acid released into a living cell will have little adverse effects. For instance, the pH and osmotic balance will not be disturbed, since the volume of the liquids confined inside the CNTs is about 10 orders of magnitude smaller than the volume of a typical living cell. Further computer simulations are required in order to estimate the temperatures at which $H_2O/CO_2$ mixtures will generate large pressures. The needed pressure values depend strongly on the type of the capping agent, which range from weakly bound polymers[41] and nanoparticles[42], to covalently bound "corks"[56], to fullerene-type CNT ends.[57] In some cases, encapsulated molecules remain inside CNTs without a cap as a result of hydrophobic and π-π stacking interactions.[16, 58] By changing the CNT diameter and the $H_2O/CO_2$ mixture composition one can adjust the vaporization temperature and control the drug release process.

**CONCLUSIONS**

Classical MD simulations were used to investigate the liquid/vapor phase transition of water confined inside CNTs of varying diameter. It was shown that spatial confinement results in a significant growth of the phase transition temperature and that a small temperature increase above the boiling point dramatically raises the inside pressure. It was found that the capillary theory successfully accounts for the boiling point elevation due to spatial confinement down to CNT diameters around 2 nm. Large deviations from the capillary theory are seen in smaller tubes. In most cases, the pressure and diffusion coefficient of gaseous water vary linearly with temperature, as expected from the kinetic theory of gases. The behavior of water inside the smallest (6,6) CNT notably differs from that in larger tubes. Below the phase transition temperature water confined within the (6,6) CNT forms an ice-like single file structure. Above the phase transition point a significant amount of order is still preserved. Even though the inside pressure grows linearly with temperature, as with the other tubes, the slope of the growth is smaller, and the diffusion coefficient remains constant. The liquid/vapor phase transition becomes a transition from an ice-like phase into a quasi-gaseous phase, in which the linear growth of pressure with temperature is driven by intense vibrations of water molecules with respect to each other rather than by translational motion as in regular gases.

The rapid growth of pressure inside CNTs at temperatures above the boiling point, in combination with precise control over boiling point elevation by the CNT diameter, leads to a proposal for a novel drug delivery system. The ability of CNTs to penetrate through cell membranes and to absorb light that passes through human tissues suggests that polar drug molecules can be packaged inside CNTs, delivered inside living cells and released by laser light. In our proposed application, the rapid growth of solvent pressure inside CNTs due to boiling should be used in order to remove CNT capping agents and discharge the drug. While the boiling temperature of water is too high for the optimal drug delivery temperature, and that of carbon dioxide is too low, a mixture of the two substances, widely known as carbonated water, should be capable of achieving the desired goal.

**METHODS**

The study was performed using classical atomistic MD simulation. A detailed description of such simulation requires definition of the classical force field, characterization of the simulation protocol and specification of the tools for data analysis. The composition of the systems under investigation is given in Table 1. Figure 1 shows one of the systems at the production run stage.

**Force fields.** The AMBER force field [59] was used in order to reproduce all bonded and non-bonded interactions within the CNTs. The simulated CNTs were treated as hydrophobic, flexible and non-polarizable particles with 1-4 carbon-carbon interactions switched on. This rather simple representation is sufficient for studying the solvent confinement effects.[60] It excludes electron polarization that can be expected in metallic CNTs, and in this respect, it makes little distinction between metallic and semiconducting tubes. Water molecules were represented by means of the TIP4P four-site force-field.[61] Having tested several force-fields, we found TIP4P to be the best one for reproducing the temperature of the liquid-vapor phase transition in bulk water. By design it reproduces the vaporization enthalpy of water at room temperature. As a known tradeoff, TIP4P noticeably overestimates the diffusion constant at room temperature ($3.7 \times 10^{-9}$ $m^2/s$). TIP4P performs well in describing the structure of liquid water. The later parameterizations of the TIP4P model focus on other aspects of bulk water. TIP4P/Ew and TIP4P/2005 match the temperature of maximum density and TIP4P/ice the melting temperature of water.[62] TIP4P-Ew improves on the boiling temperature under ambient pressure.[63] The models accounting for the density maximum of water, such as TIP4P/Ew and TIP4P/2005, provide a better estimate of the critical temperature compared to TIP4P. TIP4P/2005 can be considered as one of the best effective potentials of water for describing condensed phases, both liquid and solid.[64]

The electrostatic interactions between the solvent molecules were treated using the reaction-field-zero approach[65] with the cut-off radius of 1.3 nm. The Lennard-Jones (12,6) interactions were accounted for using the shifted force method. Hereby, the force switching region started at the interatomic distance of 1.1 nm and ended at 1.2 nm. The Lorenz-Berthelot combination rules[65] were used in order to determine the cross-term Lennard-Jones (12,6) parameters for the oxygen-hydrogen, oxygen-carbon and carbon-hydrogen intermolecular pairwise interactions.

**Simulation Protocol**. The MD simulations were carried out with the Gromacs 4.0.7 engine[65] using a 0.001 ps time-step and the leap-frog integration algorithm for solving the Newton's equations of motion.

The simulations involved several steps. The systems were prepared starting with an empty single-walled CNT, centered in the simulation box and surrounded with liquid water, as discussed in more detail in Refs.[22, 55]. Initially, no solvent molecules were located in the inner cavity of the CNT and on its outer surface. Constant temperature and pressure MD simulations were then performed for 5,000 ps in order to obtain the relaxed structures of these systems. The temperature was maintained at 300 K using the thermostat of Bussi et al.[66], often referred to as "V-rescale". The thermostat time variable was equal to 0.5 ps. The pressure was kept constant at 1bar using the Parrinello-Rahman barostat[67] with the time constant of 4.0 ps. Importantly, each considered CNT, Table 1, spontaneously and completely filled with water at room temperature and atmospheric pressure, including even the smallest (6,6) CNT with the inner diameter of just 0.52 nm. The calculation of the inner diameters of the CNTs was performed assuming that the atomic radius of carbon is 0.15 nm.

In order to overcome possible metastable states during the nanotube filling process, we used simulated annealing. After the initial filling stage was completed at 300 K, each MD system was gradually heated from 300 to 310 K during 500 ps and then cooled from 310 to 300 K during 500 ps. The heating and cooling was repeated ten times. No major changes in the potential energy were observed, indicating that a proper stable state was reached.

Before the second MD stage, all water molecules that still remained outside the nanotube, were removed. In order to represent a realistic intermolecular interaction potential of the confined water molecules, two bigger armchair CNTs were added to each system in such a way that they together formed a three-walled CNT with the same inner diameter as at the first stage. Multi-walled CNTs were selected to simplify the simulation. Proper representation of the electrostatic potential requires either outer solvent or outer tubes. Since outer liquid is not of interest in this work, we chose outer tubes. CNTs are non-polar, and therefore, are easier to handle computationally. If multi-walled CNT are used in the actual application, then the outer CNT will absorb most of the light, and a smaller fraction of light energy will reach the confined solvent.[46, 47] An example of a multi-walled

CNT system is given in Figure 1. The interwall distance in the resulting multi-walled CNTs was 0.34 nm, coinciding with the experimental value for multi-walled CNTs. The chiralities of the outer CNTs were ($n+5,n+5$), where $n$ is the chirality index of the corresponding inner tube. The resulting configurations were additionally equilibrated in the constant temperature, constant volume (NVT) ensemble during 1 000 ps.

**Data Analysis**. In order to estimate the pressure acting along the CNT axis, we used the following definition,

$$\mathbf{P} = 2 \cdot (\mathbf{E}_{kin} - \mathbf{G})/V, \qquad (1)$$

where V is the system volume, whereas $\mathbf{P}$, $\mathbf{E}_{kin}$ and $\mathbf{G}$ are the tensors of pressure, kinetic energy and virial, respectively. The virial tensor, $\mathbf{G}$, is defined as,

$$\mathbf{G} = -\frac{1}{2}\sum_{i<j}\mathbf{r}_{ij} \otimes \mathbf{F}_{ij}, \qquad (2)$$

where $r_{ij}$ and $F_{ij}$ are the radius and force vectors between particles $i$ and $j$. Note that the CNT atoms were excluded from the calculations by the above equations.

For isotropic systems, pressure can be derived using eq. (1) and (2) as

$$P = trace(\mathbf{P})/3. \qquad (3)$$

For anisotropic systems considered in the present work, we used only the pressure component along the CNT axis, focusing on pressure acting on the nanotube ends, *i.e.* $P = P_{zz}$.

The structure patterns of the confined water were analyzed in terms of the local density calculated along the axial direction of the nanotube, Fig. 5. All confined solvent atoms including hydrogens were used in the local densities calculations. The instantaneous atomic coordinates were saved every 1 ps during the 5 000 ps production runs.

The self-diffusion constant of water inside nanotubes is the most representative transport property of the liquid and vapor phases. It was calculated *via* the well-known Green-Kubo formula at several temperatures corresponding to both liquid and vapor phases. In order to calculate the velocity autocorrelation functions (VACFs), the instantaneous atomic velocities were saved every 0.005 ps during the 5 000 ps production runs. The integrals of VACF were computed numerically using a 2 000 grid point representation of VACFs.

## Acknowledgments

The authors are grateful to Drs. Yuriy Pereverzev and Heather Jaeger for comments on the manuscript. The research was supported by NSF grant CHE-1050405.

Figure 1

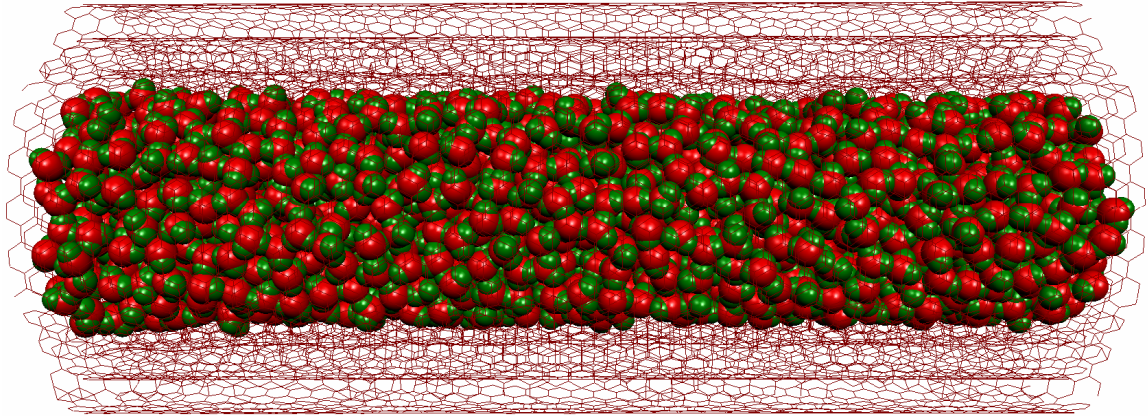

Figure 2

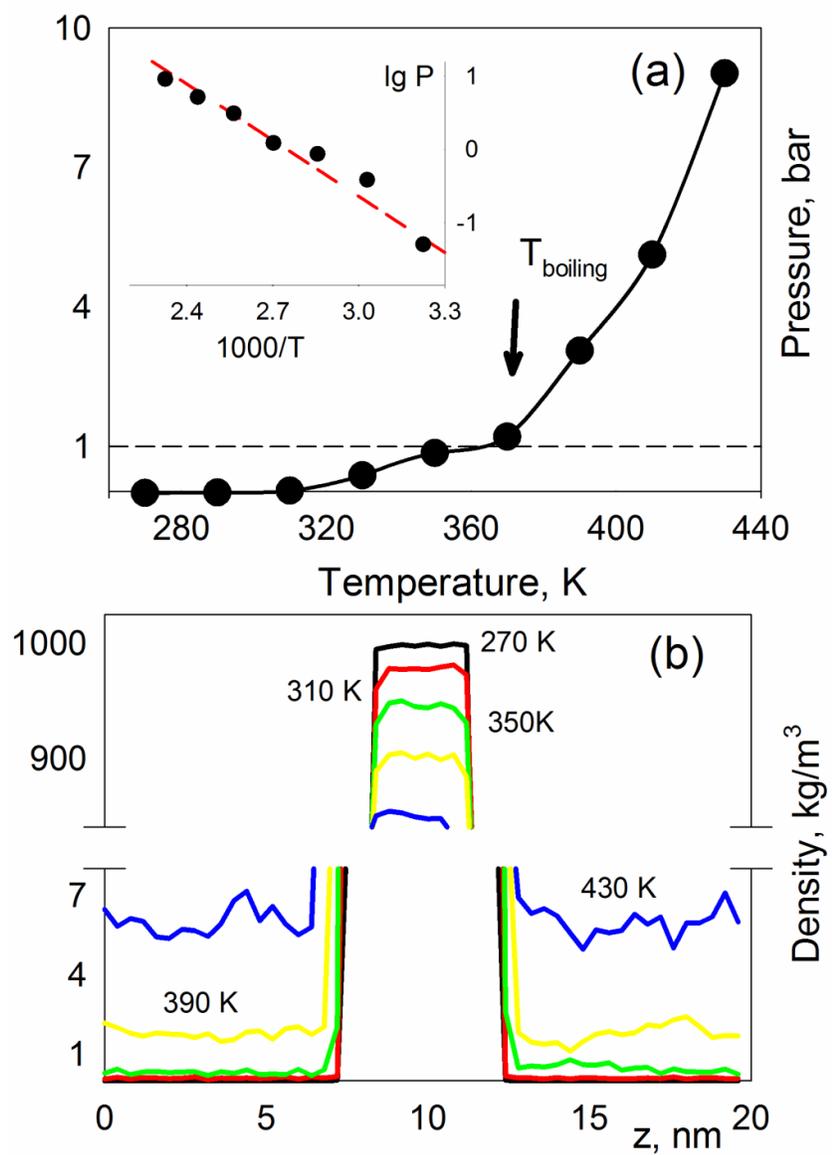

Figure 3

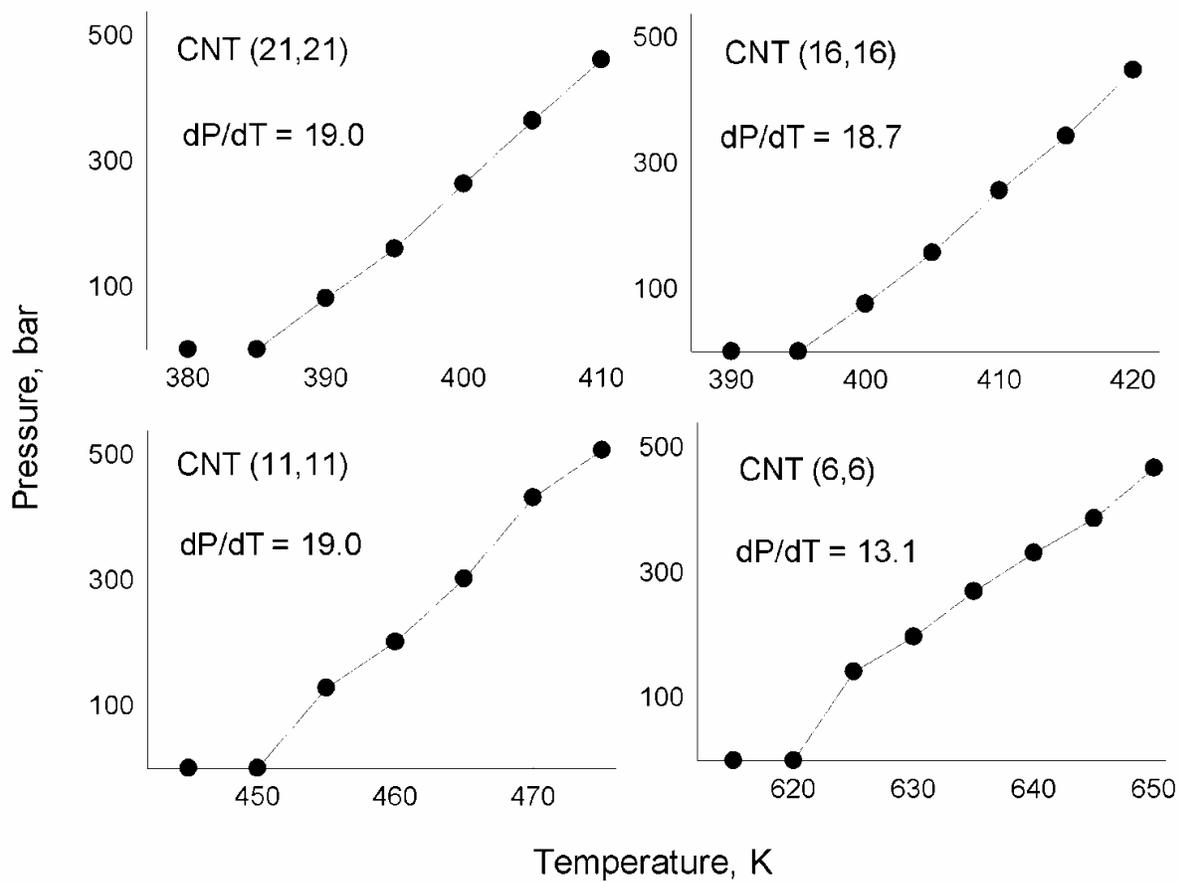

Figure 4

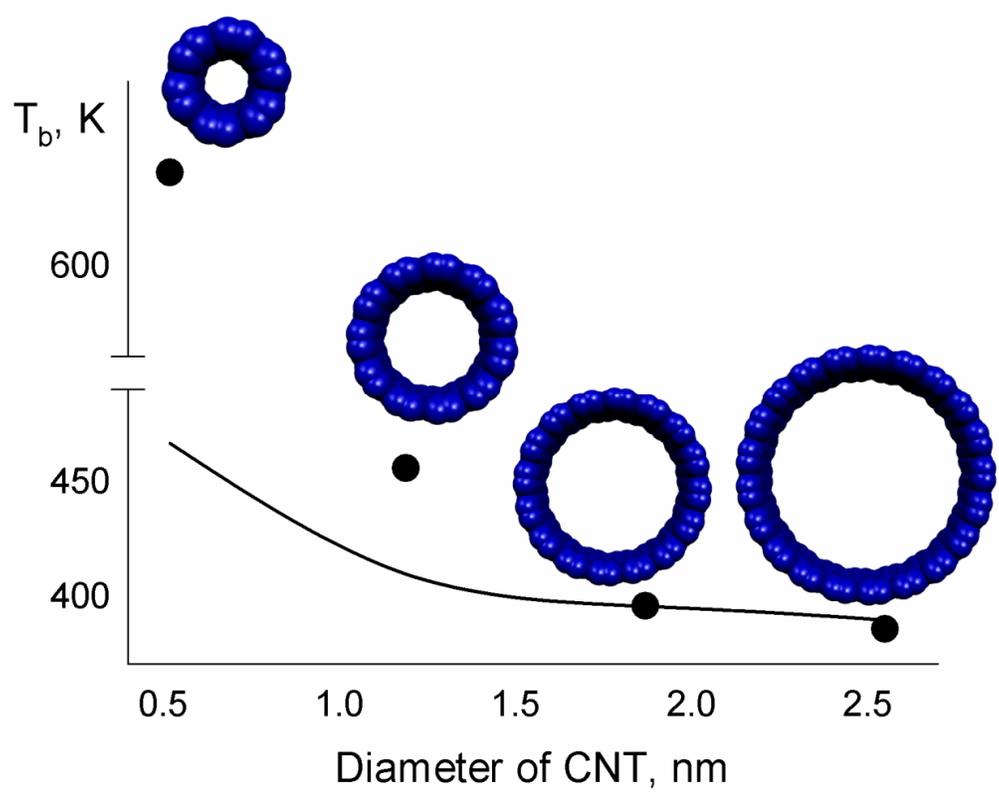

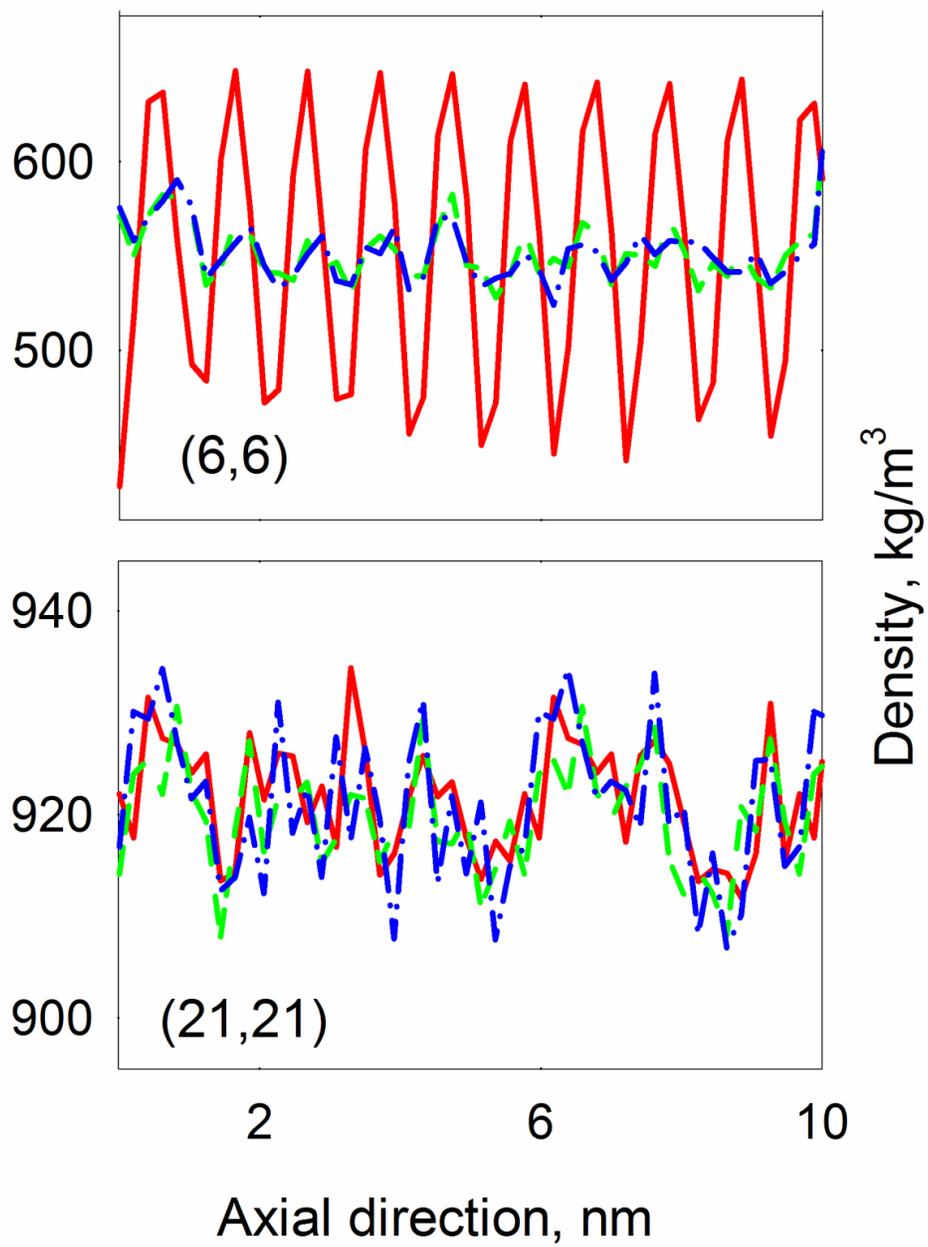

Figure 5

Figure 6

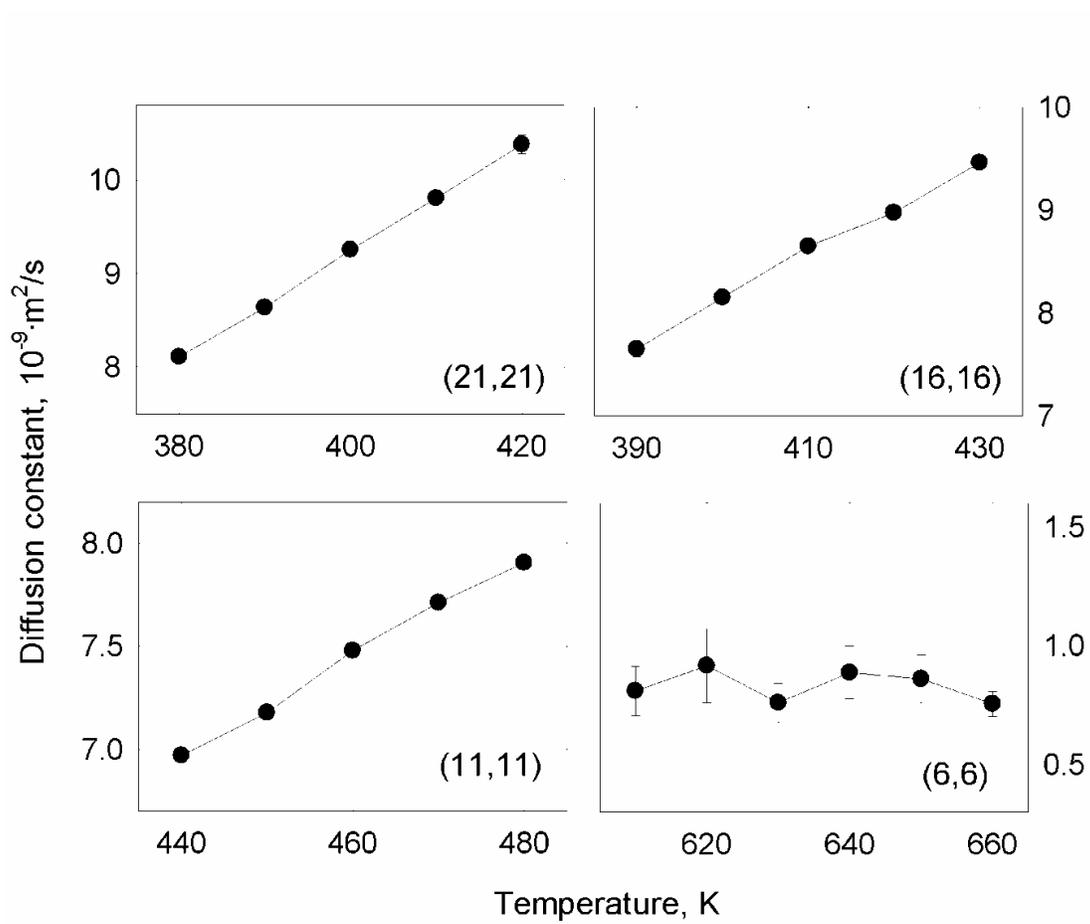

Figure 7

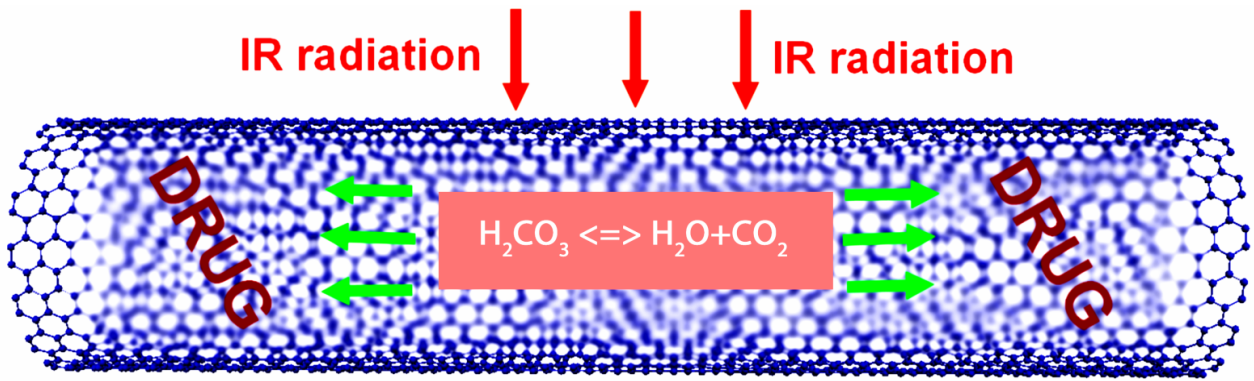

**Table 1.** Simulated molecular dynamics systems consisting of multi-walled carbon nanotubes and TIP4P water molecules. See Fig. 1 for an atomistic representation of system V.

| System | CNT[#] | $d_{CNT}$, nm (inner) | $d_{CNT}$, nm (outer) | $N_{waters}$ | $N_{waters}$(ins.) | $\rho(H_2O)^*_{ins}$, kg/m$^3$ | $T_{boiling}$, K |
|---|---|---|---|---|---|---|---|
| I | — | — | — | 1728 | — | — | 368 |
| II | (6,6) | 0.82 | 1.84 | 6110 | 40 | 642 | 625 |
| III | (11,11) | 1.49 | 2.51 | 5752 | 327 | 914 | 455 |
| IV | (16,16) | 2.17 | 3.19 | 5383 | 856 | 945 | 400 |
| V | (21,21) | 2.85 | 3.87 | 4991 | 1600 | 939 | 390 |

[#] Only the inner-most component of multi-walled CNTs is specified.
* The densities of the confined liquid were estimated assuming that the van der Waals radius of the carbon atoms of CNTs is 0.15 nm.

**Figure Captions**

**Figure 1**. (Color online) System V, see Table 1, simulated with MD includes a multi-walled CNT, composed of coaxially aligned (21,21), (26,26) and (31,31) tubes, and 1600 TIP4P water molecules inside the CNT.

**Figure 2. (a)** Saturated water vapor pressure as a function of ambient temperature derived from MD simulations of system I. The inset shows the validity of the Antoine's law (straight line) for this system. The size of the error bars is smaller than the size of the symbols. **(b)** Density profile of system I at different temperatures.

**Figure 3.** Elevation of the confined water vapor pressure with increasing temperature derived from MD simulations of systems II, III, IV and V, Table 1. The size of the error bars is smaller than the size of the symbols.

**Figure 4.** (Color online) Boiling temperature of systems II through V, Table 1, computed by MD (circles) and predicted by the capillary theory (line). The size of the error bars is smaller than the size of the symbols.

**Figure 5.** (Color online) Local density of confined water along the CNT axis. **(a)** System II at 300 K (red solid), 620 K (green dashed) and 650 K (blue dash-dotted). **(b)** System IV at 300 K (red solid), 390 K (green dashed) and 410 K (blue dash-dotted). In each case, the lowest temperature is below the boiling point, the middle temperature is just above it, and the highest temperature is well above the boiling point. The data for systems IV and V are similar to the results shown for system III.

**Figure 6.** Self-diffusion constants of water molecules as functions of temperature obtained from MD simulations for systems II through V, Table 1. The size of the error bars is smaller than the size of the symbols, with the exception of system II.

**Figure 7.** Schematic of the proposed drug delivery system.